\documentclass{IEEEtran}
\usepackage{cite}

\usepackage{amsmath,amsfonts,amssymb}
\usepackage{mathdots}
\usepackage{bbm}
\usepackage{mathtools, cuted}
\usepackage{amssymb,bm}
\usepackage{ragged2e}
\usepackage{overpic}
\usepackage{float}
\usepackage{color}
\usepackage[super]{nth}
\usepackage[normalem]{ulem}
\usepackage{graphicx}
\usepackage{textcomp}
    
\usepackage{tikz}
\usetikzlibrary{calc,positioning}
\usepackage{array}
\usepackage[caption=false,font=footnotesize]{subfig}
\usepackage[crop=pdfcrop]{pstool}
\usepackage{cancel}
\usepackage{dblfloatfix} 
\usepackage{tabularx}
\usepackage[usestackEOL]{stackengine}
\usepackage{blindtext}
\usepackage{float}
\usepackage{flushend}
\usepackage{siunitx}

\newcommand{\htext}[1]{%
	\makebox[0pt]{\Centerstack{#1}}
}

\newcommand{\vtext}[1]{%
	\makebox[0pt]{\rotatebox[origin=c]{90}{\Centerstack{#1}}}
}
\newcommand{\vtextb}[1]{%
	\makebox[0pt]{\rotatebox[origin=c]{-90}{\Centerstack{#1}}}
}

\newcommand{\tensor}[1]{\overline{\overline{#1}}}
\newcommand{\chia}[2]{\chi_{\text{#1}}^{#2}}
\newcommand{\chiat}[1]{\tensor{\chi}_{\text{#1}}}

\newcommand{\SP}[2]{S_{\text{#1}}^{\text{#2}}}

\newcommand{\TE}{\text{TE}}
\newcommand{\TM}{\text{TM}}

\definecolor{burntorange}{rgb}{0.8, 0.28, 0.0}
\definecolor{myGreen}{rgb}{0.0, 0.5, 0.0}
\definecolor{amber}{rgb}{0.8, 0.28, 0.0}
\definecolor{ceruleanblue}{rgb}{0.16, 0.28, 0.75}

\begin{document}

\title{Surface Susceptibilities as Compact Full-Wave Simulation Models of Fully-Reflective \\Volumetric Metasurfaces}

\author{Ville Tiukuvaara, \IEEEmembership{Student Member, IEEE}, Tom. J. Smy, Karim Achouri, \IEEEmembership{Member, IEEE}\\ and Shulabh Gupta, \IEEEmembership{Senior Member, IEEE}
\thanks{Ville Tiukuvaara,  Tom J. Smy, and Shulabh Gupta are with Carleton University, Ottawa, Canada (e-mail: villetiukuvaara@cmail.carleton.ca). }
\thanks{Karim Achouri is with the \'Ecole Polytechnique F\'ed\'erale de Lausanne (EPFL),
1015 Lausanne, Switzerland (e-mail: karim.achouri@epfl.ch).}
}

\maketitle

\begin{abstract}
While metasurfaces (MSs) are constructed from deeply-subwavelength unit cells, they are generally electrically-large and full-wave simulations of the complete structure are computationally expensive. Thus, to reduce this high computational cost, non-uniform MSs can be modelled as zero-thickness boundaries, with sheets of electric and magnetic polarizations related to the fields by surface susceptibilities and the generalized sheet transition conditions (GSTCs). While these two-sided boundary conditions have been extensively studied for single sheets of resonant particles, it has not been shown if they can correctly model structures where the two sides are electrically isolated, such as a fully-reflective surface. In particular, we consider in this work whether the fields scattered from a fully reflective metasurface can be correctly predicted for arbitrary field illuminations, with the source placed on either side of the surface. In the process, we also show the mapping of a PEC sheet with a dielectric cover layer to bi-anisotropic susceptibilities. Finally, we demonstrate the use of the susceptibilities as compact models for use in various simulation techniques, with an illustrative example of a parabolic reflector, for which the scattered fields are correctly computed using a integral equation (IE) based solver.
%
%
\end{abstract}



\section{Introduction}

In the past 20 years, a variety of approaches have been applied to model the behaviour of electromagnetic metamaterials, each with trade-offs in complexity, physical insight, and computational burden. MSs present an inherently ``multi-scale'' modelling problem: on one hand, they are composed of sub-wavelength scattering elements, which produce strong variations of the fields at the \textit{microscopic} level, while on the other hand the complete MSs are generally electrically large. Thus, the most rigorous method---full-wave numerical simulations---is computationally expensive. For a uniform metasurface, the application of periodic boundaries reduces the computation region to a single unit cell, but this is not possible for a non-uniform surface where the complete structure must be modelled. Doing so provides the fields at the microscopic level---which may provide physical insight---but it is not efficient for an iterative design flow due to the large simulation model.

For this reason, equivalent models have generally been used for design, which approximate the structure as a zero-thickness boundary \cite{holloway_overview_2012}. One possibility is the impedance boundary conditions (IBCs), which model the metasurface as a sheet of electric and magnetic currents, related to the tangential fields \cite{liu_generalized_2019,fong_scalar_2010,minatti_modulated_2015,monti_surface_2020}. These can be useful for multi-layer structures such as stacks of metallic patterned layers between dielectric layers, since they can be cascaded akin to elements on a transmission line \cite{wang_independent_2020}. However, a disadvantage of the IBCs is that the impedances depend on the angle of incidence. Thus, there is no unique set of impedances which truly characterizes the MS, independent of the angle of incidence. Furthermore, IBCs do not take into account normal polarization currents that may be induced; e.g., in planar split ring resonators.

Alternatively, surface susceptibilities ($\chiat{}$) can be used to represent the surface, in conjunction to the generalized sheet transition conditions (GSTCs) which provide boundary conditions on the tangential (and normal) components of the fields on either side of the surface \cite{kuester_averaged_2003,holloway_discussion_2009,albooyeh_electromagnetic_2015,achouri_general_2015,achouri_angular_2020}. The susceptibilities relate the \textit{acting} fields to the induced electric and magnetic polarization densities, where the acting fields have generally been defined one of the two following ways. One possibility is the Tretyakov-Simovski (TS) model, where the acting fields are the incident fields. The second is the Holloway-Kuester (HK) model, which defines the acting fields as the average of the total fields on either side.  The latter has an advantage in that resonances are easier to identify in the constitutive parameters (susceptibilities) \cite{albooyeh_electromagnetic_2016} and so we use the HK approach in this work. These provide true constitutive parameters that can predict the scattered fields, regardless of the incident field\footnote{The only dependence involved should be the frequency.}, provided that the susceptibilities are correctly selected and extracted. 
Consequently, the GSTCs have been implemented in a number of simulation techniques, such as finite difference methods \cite{vahabzadeh_simulation_2016,vahabzadeh_computational_2018}, the finite element method \cite{sandeep_finite-element_2017}, and integral-equation methods \cite{smy_ie-gstc_2021}, which provide efficient simulations of electrically large---and possibly curvilinear---MSs and their coupling to other scattering objects.

However, the GSTCs as they are generally written as in \cite{kuester_averaged_2003,holloway_discussion_2009,albooyeh_electromagnetic_2015,achouri_general_2015,achouri_angular_2020}, have only been shown to rigorously model structures which are composed of a single layer. Generally, these are metasurfaces composed of an array of resonators  \cite{liu_generalized_2019,kuester_averaged_2003}, while the mapping for a single dielectric layer has also been shown \cite{holloway_characterizing_2011,bhobe_derivation_2003,dehmollaian_limitations_2020}. On the other hand, one may wonder: is it possible to apply the GSTCs and the HK model to model a cascaded structure, such as dielectric layer, backed by a PEC ground-plane? While this has been considered using \textit{single-sided} boundary conditions, which involve the fields which interact with the slab and reflect from the ground-plane \cite{senior_approximate_1995,holloway_impedance-type_2000,holloway_equivalent_2000}, in this paper, we consider whether the \textit{two-sided} GSTCs can be used to model such a structure. The model should have asymmetrical behaviour, behaving as a PEC with a cover layer on one side, and a PEC on the other side. The two sides are independent (electrically isolated), so it is not immediately clear that the HK model---which uses the average of the total fields as the acting fields---should work. Surprisingly, we will show that it is indeed possible with some limitations.

The paper is structured as follows. In Section~\ref{Sec:GSTCs}, we provide the GSTCs and simplify the susceptibility tensors for the problem at hand; these are used to derive expressions for the S-parameters of a uniform surface. Next, Section~\ref{Sec:Models} gradually builds susceptibility models in increasing complexity: a PEC sheet, dielectric slab, a dielectric slab with a ground-plane, and finally the aforementioned reflective metasurface. In doing so, we use an accessible approach with notation for the GSTCs and susceptibilities that have been used in recent literature, and show the limitations of the models. In Section~\ref{Sec:Reflector}, a reflective unit cell is used to design and simulate a parabolic reflector, and a comparison is made to a full-wave simulation to show the accuracy of the susceptibility model for non-uniform metasurfaces. Finally, we conclude in Section~\ref{Sec:Conclusion}.


\section{Generalized Sheet Transition Conditions (GSTCs) and Surface Susceptibilities}\label{Sec:GSTCs}





We will use the formulation for surface susceptibilities as presented in \cite{achouri_general_2015}. Briefly, the boundary conditions at the metasurface, called the generalized surface transfer conditions (GSTCs), are
\begin{subequations}
\begin{gather}
   \hat{\mathbf{n}} \times \Delta \mathbf{H}=j \omega \mathbf{P}_\parallel-\hat{\mathbf{n}} \times \nabla M_{z}\label{eq:GSTC-dH}\\
    \hat{\mathbf{n}} \times \Delta \mathbf{E}=-j \omega \mu_{0} \mathbf{M}_\parallel-\epsilon_{0}^{-1} \hat{\mathbf{n}} \times \nabla P_{z}\label{eq:GSTC-dE}
\end{gather}\label{Eq:GSTC}
\end{subequations}
with $\Delta\phi=\phi_\text{t}-(\phi_\text{i}+\phi_\text{r})$ being the difference in fields across the boundary ($\phi\in\{\mathbf{E},\mathbf{H}\}$), and we define $\hat{\mathbf{n}}=\pm \hat{\mathbf{z}}$ as being the surface normal in the direction of incidence, directed from the side on which the incident field is present to the transmission side. 
$\textbf{P}$ and $\textbf{M}$ are the electric and magnetic surface polarization densities, respectively, with $\parallel$ denoting the projection to the boundary, while $z$ is the normal part. 

Meanwhile, the polarization densities are related to the averaged electric fields by the constitutive relations (HK model)
\begin{subequations}
\begin{align}
    \mathbf{P} =\epsilon_{0} \chiat{ee} \cdot \mathbf{E}_{\mathrm{av}}+\epsilon_{0} \eta_{0} \chiat{em} \cdot \mathbf{H}_{\mathrm{av}}\\
    \mathbf{M}=\chiat{mm} \cdot \mathbf{H}_{\mathrm{av}}+\eta_{0}^{-1} \chiat{me} \cdot \mathbf{E}_{\mathrm{av}}
\end{align}\label{Eq:constitutive}
\end{subequations}
with $\phi_\text{av}=\frac{1}{2}(\phi_\text{t}+\phi_\text{i}+\phi_\text{r})$. There are four sets of tensors, $\chiat{}$, for a total of 36 constitutive parameters. Many of these components can be eliminated or simplified due to reciprocity, symmetry, or energy conservation, depending on the particular surface \cite{achouri_angular_2020,achouri_fundamental_2021}. In this work, we will consider surfaces which involve no polarization conversion, and which are reciprocal. This simplifies the tensors to\footnote{Note that the susceptibility components $\chia{em}{xz}$, $\chia{em}{yz}$, $\chia{em}{zx}$ and $\chia{em}{zy}$ do not contribute in the prescribed problem and are thus set to zero in~\eqref{Eq:chi-tensors}.}
\begin{subequations}
	\begin{align}
		\chiat{ee} &= \begin{pmatrix}
            \chia{ee}{xx} & 0 & 0 \\
	        0 & \chia{ee}{yy} & 0 \\
	        0 & 0 & \chia{ee}{zz} \\
	    \end{pmatrix}
	    &
	    \chiat{mm} &= \begin{pmatrix}
	        \chia{mm}{xx} & 0 & 0 \\
	        0 & \chia{mm}{yy} & 0 \\
	        0 & 0 & \chia{mm}{zz} \\
	    \end{pmatrix}
	    \\
	    \chiat{em} &= 
	    \begin{pmatrix}
	        0 & \chia{em}{xy} & 0 \\
	        \chia{em}{yx} & 0 & 0 \\
	        0 & 0 & 0 \\
	    \end{pmatrix}
	    &
	     \chiat{me} &= -\chiat{em}^T\label{eq:chi-emme}
	\end{align}\label{Eq:chi-tensors}
\end{subequations}
With the 8 unique terms retained in these selected tensors, no assumptions have been made regarding energy conservation, and there is a possibility for omega-type bianisotropy \cite{asadchy_bianisotropic_2018}.

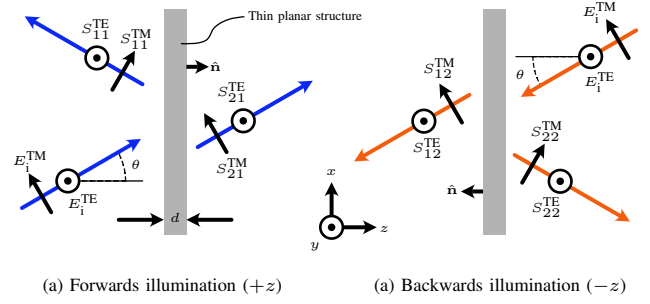
\begin{figure}
	\begin{overpic}[scale=1,grid=false,trim={0 -0.2cm 0 -0.3cm},clip]{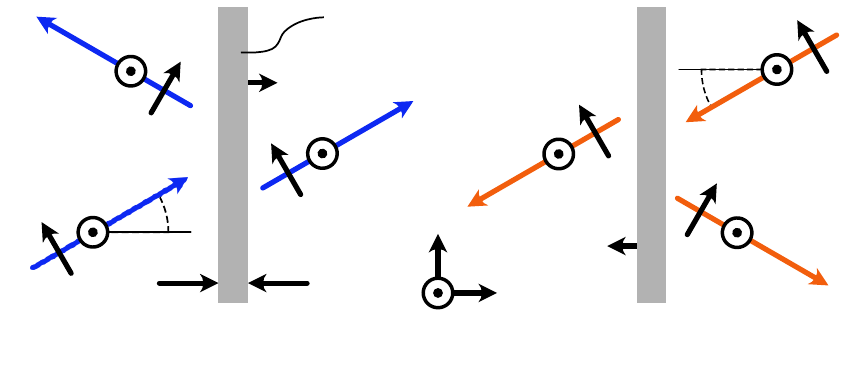}
			\put(48,8.5){\htext{\tiny $y$}}
			\put(58.5,11){\htext{\tiny $z$}}
			\put(50.5,19.5){\htext{\tiny $x$}}
			\put(26.9,12.5){\htext{\tiny $d$}}
			\put(38,43){\tiny Thin planar structure}
			\put(21,39.5){\htext{\tiny $S_{11}^\text{TM}$}}
			\put(15,41){\htext{\tiny $S_{11}^\text{TE}$}}
			\put(35.5,19.9){\htext{\tiny $S_{21}^\text{TM}$}}
			\put(35.5,31.5){\htext{\tiny $S_{21}^\text{TE}$}}
			\put(5,21){\htext{\tiny $E_\text{i}^\text{TM}$}}
			\put(13,14.5){\htext{\tiny $E_\text{i}^\text{TE}$}}
			\put(66.8,35){\htext{\tiny $S_{12}^\text{TM}$}}
			\put(65,23){\htext{\tiny $S_{12}^\text{TE}$}}
			\put(83,25.5){\htext{\tiny $S_{22}^\text{TM}$}}
			\put(83,14){\htext{\tiny $S_{22}^\text{TE}$}}
			\put(91,43.5){\htext{\tiny $E_\text{i}^\text{TM}$}}
			\put(91,33){\htext{\tiny $E_\text{i}^\text{TE}$}}
			\put(21,20.5){\htext{\tiny $\theta$}}
			\put(79,34){\htext{\tiny $\theta$}}
            \put(25,2){\htext{\scriptsize (a) Forwards illumination ($+z$)}}
            \put(76,2){\htext{\scriptsize (a) Backwards illumination ($-z$)}}
            \put(33,35.3){\htext{\tiny $\hat{\mathbf{n}}$}}
            \put(68.7,16.7){\htext{\tiny $\hat{\mathbf{n}}$}}
    \end{overpic}
	\caption{A depiction of the electric field orientations for TE and TM fields, with forwards and backwards plane wave illumination of a planar structure (e.g. a metasurface). Propagation is in the $x-z$ plane, with the metasurface in the $x-y$ plane.}\label{Fig:Problem}
\end{figure}

Since the susceptibilities in \eqref{Eq:chi-tensors} do not convert polarization, TE and TM illuminations can be considered separately, as depicted in Figure ~\ref{Fig:Problem}. With the periodicity of the surface being subwavelength, no higher-order diffraction orders are generated \cite{tiukuvaara_floquet_2021} and under oblique plane wave illumination at $\theta$, there will be reflected and transmitted plane waves at the same angle, following standard Snell's laws. The ``ports'' on the left and right sides are denoted 1 and 2, respectively, such that e.g. $S_{21}^{\{\text{TE},\text{TM}\}}$ denotes transmission in the forwards direction and $S_{12}^{\{\text{TE},\text{TM}\}}$ in the backwards direction. Substituting the expressions for the plane waves along with \eqref{Eq:chi-tensors} into \eqref{Eq:GSTC} and \eqref{Eq:constitutive}, and solving the resulting system of equations for the S-parameters, we obtain
\begin{subequations}
    \begin{align}
        S_{\{11,22\}}^{\TE}(\theta) &= \frac{2jk}{\xi_\TE} \left(\zeta_\TE\mp 2\chia{em}{yx}\cos\theta-\chia{mm}{xx}\cos^2\theta\right)\\
        S_{\{21,12\}}^{\TE}(\theta) &= \frac{\cos\theta}{\xi_\TE} \left(4+k^2\left[ \left(\chia{em}{yx}\right)^2+\chia{mm}{xx}\zeta_\TE\right]\right)\\
        S_{\{11,22\}}^{\TM}(\theta) &= \frac{2jk}{\xi_\TM} \left(\zeta_\TM\mp 2\chia{em}{xy}\cos\theta-\chia{ee}{xx}\cos^2\theta\right)\\
        S_{\{21,12\}}^{\TM}(\theta) &= \frac{\cos\theta}{\xi_\TM} \left(4+k^2\left[ \left(\chia{em}{xy}\right)^2+\chia{ee}{xx}\zeta_\TM\right]\right)
    \end{align}
where the top and bottom signs ($\mp$) are taken for $11/21$ and $22/12$, respectively, and
\begin{multline}
    \xi_{\{\TE,\TM\}}=-4\cos\theta-2jk\left(\zeta_{\{\TE,\TM\}}+\chi_{\{\text{mm},\text{ee}\}}^{xx}\cos^2\theta\right)\\
    +k^2\cos\theta\left(\zeta_{\{\TE,\TM\}}\chi_{\{\text{mm},\text{ee}\}}^{xx}+\left(\chia{em}{\{yx,xy\}}\right)^2\right)
\end{multline}
\begin{gather}
    \zeta_{\{\TE,\TM\}} = \chi_{\{\text{ee},\text{mm}\}}^{yy}+\chi_{\{\text{mm},\text{ee}\}}^{zz}\sin^2\theta
\end{gather}\label{Eq:Sparams}
\end{subequations}

We note in particular that the bianisotropic terms $\chia{em}{\{yx,xy\}}$ [and implicitly terms in $\chiat{me}$ following \eqref{eq:chi-emme}] lead to an asymmetry in reflection when the direction of illumination is flipped, as noted in \cite{achouri_angular_2020}. If $\chia{em}{yx}=0$ or $\chia{em}{xy}=0$, then $S_{11}^{\TE}=S_{22}^{\TE}$ or $S_{11}^{\TM}=S_{22}^{\TM}$, respectively. At the same time, $S_{21}^{\TE}=S_{12}^{\TE}$ and $S_{21}^{\TM}=S_{12}^{\TM}$ always hold true, which is a result of enforcing reciprocity in \eqref{Eq:chi-tensors}. Finally, the fact that $S_{\{12,21\}}^{\{\text{TE,TM}\}}(\theta)=S_{\{12,21\}}^{\{\text{TE,TM}\}}(-\theta)$ implies that the metasurface leads to angular symmetric scattering \cite{achouri_angular_2020}.

\section{Surface Susceptibility Models}\label{Sec:Models}

We will use \eqref{Eq:Sparams} to determine the susceptibility models for several simple structures, progressively building up to a reflective metasurface, for which the susceptibilities correctly model the behaviour with illumination from either side. We use a pedagogical approach, starting with simple structures such as a PEC sheet and a dielectric sheet to motivate the susceptibility terms in \eqref{Eq:chi-tensors} and understand their role for the final reflective metasurface.

\subsection{PEC Sheet} \label{Sec:PECSheet}

Firstly, consider a perfect electric conductor (PEC) sheet, with $S_{\{21,12\}}^{\{\TE,\TM\}}=0$ and $S_{\{11,22\}}^{\{\TE,\TM\}}=-1$. Given the reflection symmetry, we must have $\chia{em}{\{yx,xy\}}=0$. A PEC is the limiting case of a conductor with infinite conductivity ($\sigma\rightarrow\infty$), which in fact, corresponds to a limiting case of $\chia{ee}{xx}$ and $\chia{ee}{yy}$. By eliminating all other susceptibility terms from \eqref{Eq:Sparams} and taking a limit, we find\footnote{We consider that in a conductive material $\epsilon = \epsilon' - j\sigma/\omega$ and that $\epsilon=\epsilon_0(1+\chi)$. So, in a PEC, we have that $\epsilon \rightarrow -j\infty$ and thus $\chi \rightarrow -j\infty$.}
\begin{subequations}
    \begin{gather}
        \lim_{\chia{ee}{yy}\rightarrow-j\infty}S_{\{11,22\}}^{\TE}=
        \lim_{\chia{ee}{yy}\rightarrow-j\infty}\left(\frac{k}{\frac{2j\cos\theta}{\chia{ee}{yy}}-k}\right) = -1\\
        \lim_{\chia{ee}{yy}\rightarrow-j\infty}S_{\{21,12\}}^{\TE}=
        \lim_{\chia{ee}{yy}\rightarrow-j\infty}\left(\frac{2\cos\theta}{2\cos\theta+jk\chia{ee}{yy}}\right) = 0\\
        \lim_{\chia{ee}{xx}\rightarrow-j\infty}S_{\{11,22\}}^{\TM}=
        \lim_{\chia{ee}{xx}\rightarrow-j\infty}\left(\frac{k\cos\theta}{\frac{2j}{\chia{ee}{xx}}-k\cos\theta}\right) = -1\\
        \lim_{\chia{ee}{xx}\rightarrow-j\infty}S_{\{21,12\}}^{\TM}=
        \lim_{\chia{ee}{xx}\rightarrow-j\infty}\left(\frac{2j}{2j-k\chia{ee}{xx}\cos\theta}\right) = 0
    \end{gather}
\end{subequations}
Thus, $\chia{ee}{\{xx,yy\}}\rightarrow-j\infty$ precisely models a PEC sheet.\footnote{Similarly, $\chia{mm}{\{xx,yy\}}\rightarrow-j\infty$ would be a PMC.} However, this is not suitable for numerical simulation, since it requires a limit. 

Now, we consider if there is another susceptibility model, having finite susceptibilities, which could be used instead. To this end, we return to considering all 8 terms in \eqref{Eq:chi-tensors}. Of course, including the bianisotropic terms produces asymmetry with forwards/backwards illumination, and so here we seek a possible compromise that may not rigorously model a PEC but is useful for numerical implementation. We will enforce $S_{\{12,21\}}^{\{\TE,\TM\}}(\theta)=0$ and $S_{11}^{\{\TE,\TM\}}(\theta)=-1$ (PEC with forwards illumination); these provide 6 equations from \eqref{Eq:Sparams} while $S_{22}^{\{\TE,\TM\}}(\theta)$ is not enforced. To reduce the number of unknown terms from 8 to match the number of equations, we consider symmetry. There is rotational symmetry around the $z$ axis, i.e. isotropy, and so the susceptibilities should remain unchanged with this rotation. That is
\begin{align}
	\chiat{}\stackrel{?}{=}\overline{\overline{R}}_z(\phi)\cdot\chiat{}\cdot \overline{\overline{R}}_z(\phi)^T\label{Eq:Rotation}
\end{align}
should hold true, where $\overline{\overline{R}}_z(\phi)$ is the transformation matrix which rotates by $\phi$ about the $z$ axis \cite{achouri_fundamental_2021}. This requirement means $\chi_{\{\text{ee,mm}\}}^{xx}=\chi_{\{\text{ee,mm}\}}^{yy}$ leaving 6 unknown susceptibilities.
 Solving the resulting system of equations yields the non-zero terms $\chia{em}{yx}=+2j/k$ and $\chia{em}{xy}=-2j/k$. Note that these terms substituted into the bianisotropic tensors (\ref{Eq:chi-tensors}b) also satisfy \eqref{Eq:Rotation}.
 

Now, what happens for backwards illumination? Using  (\ref{Eq:Sparams}a,c) we find $S_{22}^{\{\TE,\TM\}}(\theta)=+1$. Thus, the model appears as a PEC with forwards illumination and a PMC with backwards illumination, independent of the angle of incidence. If $\chi_\text{em}^{\{xy,yx\}}$ are negated, the structure is effectively reflected in the $x-y$ plane. The angular independence with these purely bi-anisotropic susceptibilities was studied in \cite{achouri_electromagnetic_2021}, where it was also noted that such a surface is only possible to fabricate with a physical unit cell at a single frequency, in the limit of zero loss. For this work, however, the inability to physically represent a PEC sheet (with correct behaviour on both sides) does not end up being critical in our our subsequent objective, when we later consider a dielectric cover layer added to the PMC side.




\subsection{Dielectric Sheet}\label{Sec:Slab}

Building towards a resonator on top of a ground-plane with a cover layer, we next consider an isolated sheet of uniform permittivity $\epsilon_r=(\epsilon_r'-j\epsilon_r'')$, having an imaginary part allowing for loss, with a thickness $d$. This mapping was considered in \cite{dehmollaian_limitations_2020}, where only the tangential $\{\text{ee},\text{mm}\}$ susceptibilities were used and earlier in \cite{bhobe_derivation_2003} where  the normal components ($\chi_{\{\text{ee},\text{mm}\}}^{zz}$) were included. We will consider the extraction here for completeness, using a simple and accessible approach.

The analytical reflection and transmission through a dielectric slab, are well-known, given by
\begin{multline}
	\begin{bmatrix}
		1\\ S_{11}^\text{a}(\theta)
	\end{bmatrix}
	=
	\underbrace{\frac{1}{\tau_1^\text{a}(\theta)}
	\begin{bmatrix}
		1 & \rho_1^\text{a}(\theta)\\
		\rho_1^\text{a}(\theta) & 1
	\end{bmatrix}}_{\substack{\text{air-dielectric interface}}}
	\cdot
	\underbrace{\begin{bmatrix}
		e^{j\phi} & 0\\
		0 & e^{-j\phi}
	\end{bmatrix}}_{\substack{\text{propagation}\\\text{in dielectric}}}\\
	\cdot
	\underbrace{\frac{1}{\tau_2^\text{a}(\theta)}
	\begin{bmatrix}
		1 & \rho_2^\text{a}(\theta)\\
		\rho_2^\text{a}(\theta) & 1
	\end{bmatrix}}_{\substack{\text{dielectric-air interface}}}
	\cdot
	\begin{bmatrix}
		S_{21}^\text{a}(\theta)\\0
	\end{bmatrix}\label{Eq:SlabCascade}
\end{multline}
with $a\in\{\text{TE},\text{TM}\}$, $\phi=kd\sqrt{\epsilon_r}\cos\theta$, and where $\rho_{\{1,2\}}^\text{a}(\theta)$ and $\tau_{\{1,2\}}^\text{a}(\theta)$ are the Fresnel coefficients for oblique incidence at the first and second interfaces, dependent on the polarization (TE/TM) \cite{orfanidis_electromagnetic_2016}. The matrix equation can be solved to provide a total of four expressions for $S_{\{11,21\}}^{\{\text{TE,TM}\}}(\theta)$.

With $\theta=\SI{0}{\degree}$, the normal susceptibility terms in (\ref{Eq:Sparams}f) are eliminated, and we set  $\chi_\text{em}^{\{yx,xy\}}=0$ so that there is symmetry between forwards and backwards illumination, we can substitute $S_{\{11,21\}}^{\{\text{TE,TM}\}}(0)$ from \eqref{Eq:SlabCascade} into \eqref{Eq:Sparams} to have four equations with the solution
\begin{subequations}
	\begin{align}
		\chia{ee}{xx} = \chia{ee}{yy} &=\frac{2 \sqrt{\epsilon _r} \tan \left(\frac{kd\sqrt{\epsilon _r}}{2} \right)}{k}\\
        \chia{mm}{xx} = \chia{mm}{yy} &=\frac{2 \tan \left(\frac{kd\sqrt{\epsilon _r}}{2}\right)}{k \sqrt{ \epsilon _r}}
	\end{align}
Repeating a similar procedure for $\theta\neq \SI{0}{\degree}$, the normal susceptibility terms are found to be
	\begin{align}
		\chia{ee}{zz} &= \csc^2\theta \left(
		\frac{2\gamma\tan\left(\frac{kd\gamma}{2} \right)}{k\sqrt{\epsilon_r}}-\chia{mm}{xx}
		\right)\\
		\chia{mm}{zz} &= \csc^2\theta \left(
		\frac{2\gamma\tan\left(\frac{kd\gamma}{2} \right)}{k}-\chia{ee}{xx}
		\right)
	\end{align}\label{Eq:SlabChi}
\end{subequations}
with $\gamma=\sqrt{\epsilon_r-\sin^2\theta}$. As is, there is an angular dependence in (\ref{Eq:SlabChi}c-d), which should not be the case, if $\chi_{\{\text{ee,mm}\}}^{zz}$ are to be true constitutive parameters, independent of the incident field \cite{achouri_angular_2020}. However, we can only expect a zero-thickness model to apply for thin slabs ($kd\ll 1$), and so we proceed by using the first few terms of the Taylor expansion around $kd=0$:
\begin{subequations}
	\begin{align}
		\chia{ee}{zz} &= -\frac{d}{\epsilon_r}-\frac{k^2d^3}{6}+\cancelto{0}{\frac{k^2d^3\sin^2\theta}{12\epsilon_r}}+ O[(kd)^4]\\
        \chia{mm}{zz} &= -d-\frac{k^2d^3}{6\epsilon_r}+\cancelto{0}{\frac{k^2d^3\sin^2\theta}{12}}+ O[(kd)^4]
	\end{align}\label{Eq:SlabChiExpansion}
\end{subequations}%
\noindent with the approximation that the third term and higher-order terms are negligible. Within this approximation, $\chi_{\{\text{ee,mm}\}}^{zz}$ only depends on the properties of the slab (at a given frequency), and can be considered constitutive parameters.

\begin{figure}
    \begin{overpic}[grid=false,trim=0cm 0cm 0cm 0cm,clip]{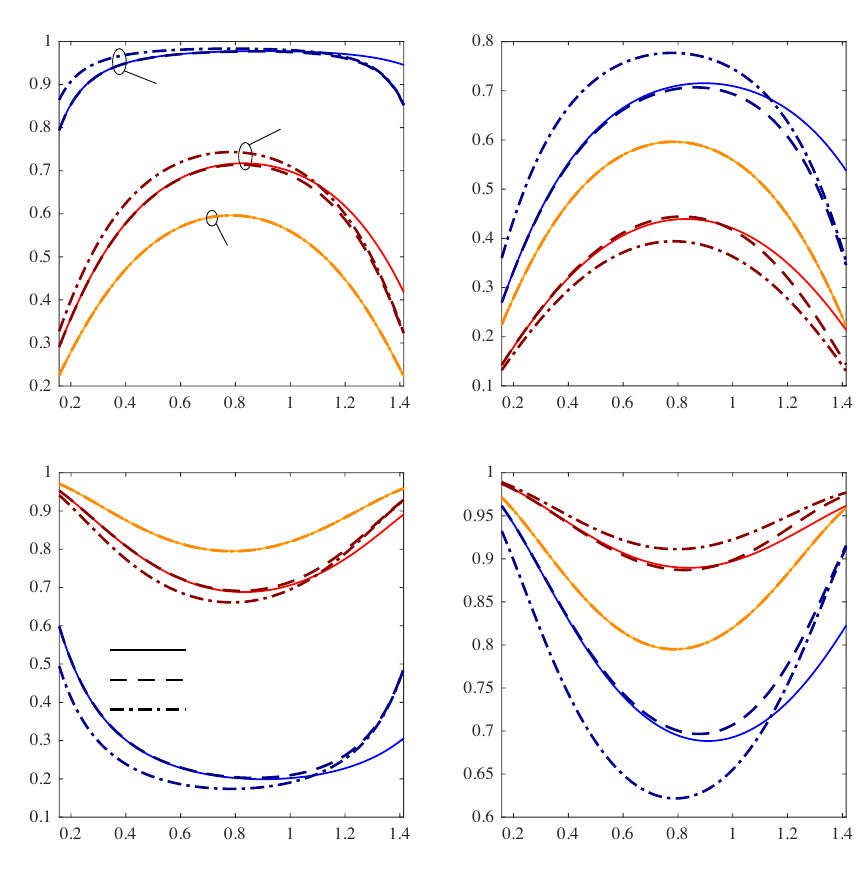}
        \put(26.5,2){\htext{\scriptsize $kd$}}
        \put(76.5,2){\htext{\scriptsize $kd$}}
        \put(26.5,51){\htext{\scriptsize $kd$}}
        \put(76.5,51){\htext{\scriptsize $kd$}}
        \put(1,26){\vtext{\scriptsize $|\SP{21}{TE}|$}}
        \put(50,26){\vtext{\scriptsize $|\SP{21}{TM}|$}}
        \put(1,75){\vtext{\scriptsize $|\SP{11}{TE}|$}}
        \put(50,75){\vtext{\scriptsize $|\SP{11}{TM}|$}}
        \put(18,89){\tiny $\theta=\SI{60}{\degree}$}
        \put(32,85){\tiny $\theta=\SI{30}{\degree}$}
        \put(25,70){\tiny $\theta=\SI{00}{\degree}$}
        \put(22,26){\tiny Analytical, \eqref{Eq:SlabCascade}}
        \put(22,22.5){\tiny $\chiat{}$, \eqref{Eq:SlabChi} \& \eqref{Eq:SlabChiExpansion}}
        \put(22,19.5){\tiny $\chiat{}$ with $\chia{}{zz}=0$}
    \end{overpic}
	\caption{The reflection and transmission through a lossy dielectric sheet ($\epsilon_r=4-j0.04$) was calculated analytically using \eqref{Eq:SlabCascade} to compare with the mapped susceptibilities \eqref{Eq:SlabChi}, with and without the normal components.}\label{Fig:SlabSparams}
\end{figure}

To validate the assumption and verify \eqref{Eq:SlabChi} and \eqref{Eq:SlabChiExpansion}, we consider a numerical example, shown in Figure~\ref{Fig:SlabSparams}. The reflection and scattering for both TE and TM illuminations is shown, for increasing thickness $kd$ and at three angles, $\theta\in\{\SI{0}{\degree},\SI{30}{\degree},\SI{60}{\degree}\}$. Firstly, we see that even when the slab is very thin ($kd=0.2$, i.e. $d\approx \lambda/30$) the normal components $\chi_{\{\text{ee,mm}\}}^{zz}$ are needed to model the scattering at oblique incidence. Secondly, the modelling is very accurate up to about $kd=0.8$ ($d\approx \lambda/8$), past which the analytical results diverge.

\begin{figure}
	\begin{overpic}[grid=false,trim=0cm 0cm 0cm 0cm,clip]{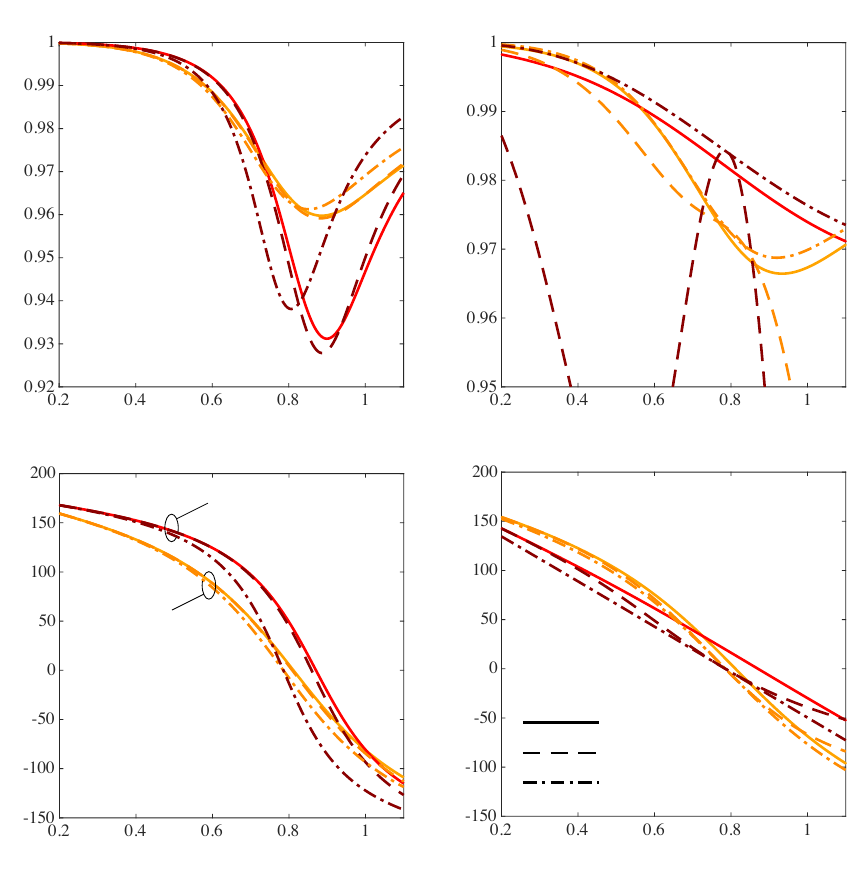}
        \put(26.5,2){\htext{\scriptsize $kd$}}
        \put(76.5,2){\htext{\scriptsize $kd$}}
        \put(26.5,51){\htext{\scriptsize $kd$}}
        \put(76.5,51){\htext{\scriptsize $kd$}}
        \put(0,26){\vtext{\scriptsize $\angle{}\SP{11}{TE}$ (degrees)}}
        \put(50,26){\vtext{\scriptsize $\angle{}\SP{11}{TM}$ (degrees)}}
        \put(0,75){\vtext{\scriptsize $|\SP{11}{TE}|$}}
        \put(50,75){\vtext{\scriptsize $|\SP{11}{TM}|$}}
        \put(24,43){\tiny $\theta=\SI{60}{\degree}$}
        \put(10,30){\tiny $\theta=\SI{30}{\degree}$}
        %
        \put(68,18){\tiny Analytical, \eqref{Eq:GroundedSlabAnalytical}}
        \put(68,14.5){\tiny $\chiat{}$, \eqref{Eq:GroundedSlabChi}}
        \put(68,11.3){\tiny $\chiat{}$ with $\chia{}{zz}=0$}
    \end{overpic}
	\caption{The reflection from a PEC sheet with a lossy dielectric cover layer ($\epsilon_r=4-j0.04$) was calculated analytically using \eqref{Eq:GroundedSlabAnalytical} to compare with the mapped susceptibilities \eqref{Eq:GroundedSlabAnalytical}, with and without the normal components. Meanwhile, $S_{22}^{\{\text{TE,TM}\}}=-1$, and $S_{12,21}^{\{\text{TE,TM}\}}=0$ (not plotted).}\label{Fig:GroundedSlabSparams}
\end{figure}

\begin{figure*}[h]
	\begin{overpic}[grid=false,trim=0cm 0cm 0cm 0cm,clip]{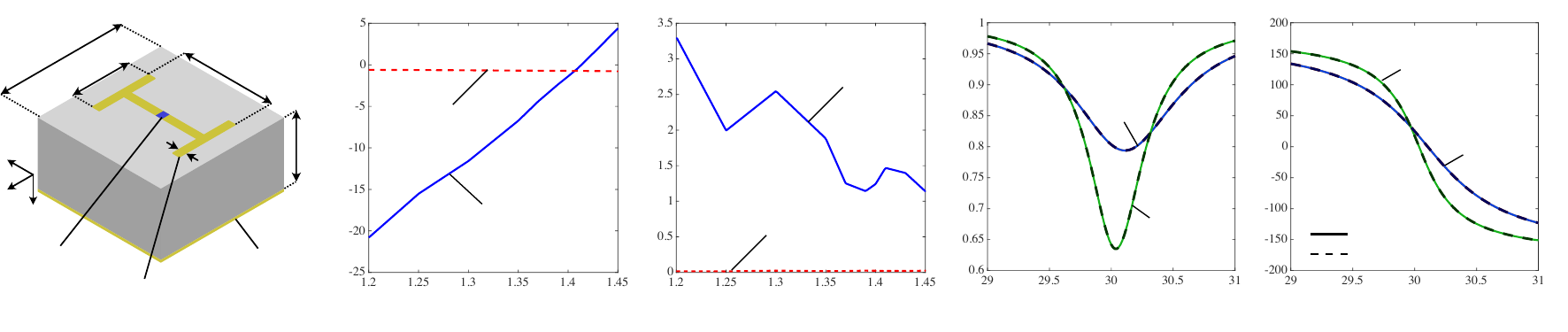}
		\put(10,0){\htext{\footnotesize (a)}}
		\put(41,0){\htext{\footnotesize (b)}}
		\put(80,0){\htext{\footnotesize (c) }}
		\put(3.9,4.3){\htext{\tiny Lumped inductor ($L$)}}
		\put(16.6,4.3){\htext{\tiny PEC ground-plane}}
		\put(9,2.5){\htext{\tiny Copper dipole}}
		\put(5.1,16.4){\htext{\tiny 0.4 mm}}
		\put(3,18){\htext{\tiny 1.0 mm}}
		\put(15.5,16.7){\htext{\tiny 0.7 mm}}
		\put(19.3,11.4){\vtext{\tiny $d=508$ $\mu$m}}
		\put(13.8,9.8){\htext{\tiny 50 $\mu$m}}
		\put(31.5,2){\htext{\tiny $L$ (nH)}}
		\put(21.5,11){\vtext{\tiny $\Re\{\chia{ee}{yy}\}$, $\Im\{\chia{ee}{yy}\}$ ($\times 10^{-3}$)}}
		\put(31.5,7){\htext{\tiny $\Re$}}
		\put(28,14){\htext{\tiny $\Im$}}
		\put(51,2){\htext{\tiny $L$ (nH)}}
		\put(41,11){\vtext{\tiny $\Re\{\chia{mm}{zz}\}$, $\Im\{\chia{mm}{zz}\}$ ($\times 10^{-3}$)}}
		\put(54.5,16){\htext{\tiny $\Re$}}
		\put(49.5,6.5){\htext{\tiny $\Im$}}
		\put(71,2){\htext{\tiny Frequency (GHz)}}
		\put(60.3,11){\vtext{\tiny $|\SP{11}{TE}|$}}
		\put(69.8,13.8){\tiny $\theta=\SI{0}{\degree}$}
		\put(73.7,6.5){\tiny $\theta=\SI{60}{\degree}$}
		\put(90.5,2){\htext{\tiny Frequency (GHz)}}
		\put(80,11){\vtext{\tiny $\angle{}\SP{11}{TE}$ (degrees)}}
		\put(93.5,11.3){\tiny $\theta=\SI{0}{\degree}$}
		\put(89.8,16.6){\tiny $\theta=\SI{60}{\degree}$}
		\put(86.5,6){\tiny HFSS}
		\put(86.5,4.6){\tiny $\chiat{}$, \eqref{Eq:UnitCellChi}}
		\put(0,8.5){\htext{\tiny $x$}}
		\put(0,11){\htext{\tiny $y$}}
		\put(2,7.5){\htext{\tiny $z$}}
	\end{overpic}
	\caption{A deeply-subwavelength reflective unit cell was designed using an electric dipole, loaded with a lumped inductor ($L$) on top of a Rogers RO4003C substrate ($\epsilon_r=3.55$, $\tan\delta_d=0.0027$) that is on a PEC ground-plane. (a) Unit cell model. (b) $\chiat{}$ extraction at \SI{30}{GHz} using \eqref{Eq:UnitCellChi}, as a function of the loading inductance $L$. (c) $\SP{11}{TE}$ with $L=\SI{1.4}{nH}$ at two angles of incidence.}\label{Fig:CellExtraction}
\end{figure*}
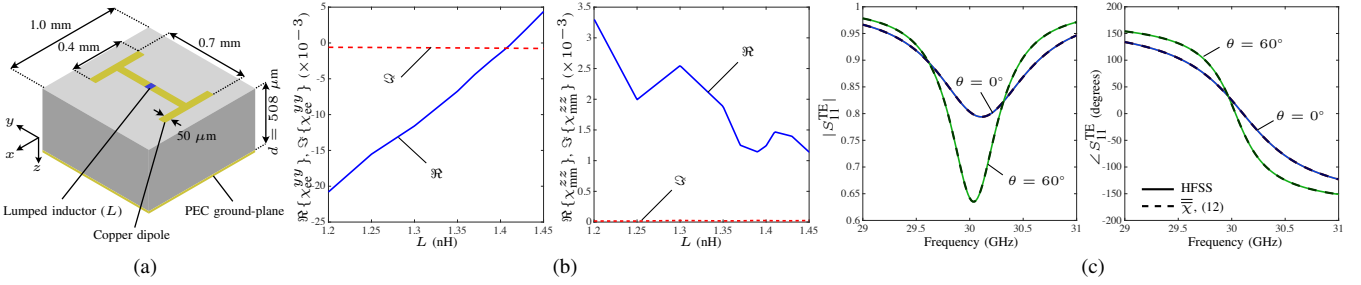

\subsection{PEC Sheet with Cover Layer}\label{Sec:GroundedSlab}


Next, consider a PEC with a dielectric cover layer on one side. In this case, the forwards and backwards illuminations clearly have non-symmetrical behaviour. With the cover layer on the left side of the PEC sheet in Figure~\ref{Fig:Problem}, the forwards illumination case is governed by
\begin{align}
	\begin{bmatrix}
        1\\S_{11}^\text{a}(\theta)
    \end{bmatrix}=
    \underbrace{\frac{1}{\tau_1(\theta)}
    \begin{bmatrix}
		1 & \rho_1^\text{a}(\theta)\\
		\rho_1^\text{a}(\theta) & 1
	\end{bmatrix}}_{\substack{\text{air-dielectric interface}}}
	\cdot
	\underbrace{\begin{bmatrix}
		e^{j\phi} & 0\\
		0 & e^{-j\phi}
	\end{bmatrix}}_{\substack{\text{propagation}\\\text{in dielectric}}}
	\cdot
    \begin{bmatrix}
        E_\text{PEC}^\text{a}\\
        -E_\text{PEC}^\text{a}
    \end{bmatrix}\label{Eq:GroundedSlabAnalytical}
\end{align}
where $E_\text{PEC}^\text{a}$ is the amplitude of the forwards-travelling wave at the PEC. Since $E_\text{PEC}^\text{a}$ is not of interest, this reduces to two expressions for $S_{11}^{\{\text{TE,TM}\}}(\theta)$, one for each polarization. We get six equations from $S_{11}^{\{\text{TE,TM}\}}(0)$, $S_{11}^{\{\text{TE,TM}\}}(\theta)$, and  $S_{21}^{\{\text{TE,TM}\}}(\theta)$ [equating \eqref{Eq:GroundedSlabAnalytical} and \eqref{Eq:Sparams}], and two more equations from  $S_{22}^{\{\text{TE,TM}\}}(0)=-1$, for a total of eight equations. Solving these,
\begin{subequations}
	\begin{gather}
		\chia{mm}{\{xx,yy\}}= 0\\
        \chia{em}{yx} = -\frac{2j}{k_0}\\
        \chia{em}{xy} = +\frac{2j}{k_0}\\
        \chia{ee}{\{xx,yy\}} =  -\frac{4  \cot \left(kd\sqrt{\epsilon _r}\right)}{\sqrt{\epsilon_r}}\\
        \chia{ee}{zz} \approx -\frac{4d}{\epsilon_r} -\frac{8 k^2d^3}{3}\\
        \chia{mm}{zz} \approx -\frac{4d}{3}-\frac{8k^2d^3\epsilon_r}{45}
	\end{gather}\label{Eq:GroundedSlabChi}
\end{subequations}
after all but the leading angle-independent terms in the Taylor expansion are retained for (\ref{Eq:GroundedSlabChi}e,f) (the complete expressions for $\chi_{\{\text{ee,mm}\}}^{zz}$ are cumbersome and not presented). It is interesting that $\chi_\text{em}^{\{yx,xy\}}$ are in fact identical to those for a sheet that behaves as a PEC on the back side ($S_{22}^{\{\text{TE,TM}\}}=-1$) and a PMC on the front side ($S_{11}^{\{\text{TE,TM}\}}=+1$), as discussed in Section~\ref{Sec:PECSheet}. However, in this case the other susceptibilities contribute such that the front does not appear as a PMC, but a PEC with a cover layer.

To check that this is is indeed the case, numerical results are shown in Figure~\ref{Fig:GroundedSlabSparams}. For TE polarization, \eqref{Eq:GroundedSlabChi} provides a good match for the forwards reflection past $kd>1$, in comparison to the analytical expression \eqref{Eq:GroundedSlabAnalytical}; the resonance at $kd\approx 0.9$ is correctly predicted. For TM polarization, there is reasonable agreement if $\chia{ee}{zz}$ is neglected, but including causes a large disagreement. Thus, it seems that the truncation of the Taylor series in (\ref{Eq:GroundedSlabChi}e) is a poor approximation, and this term should not be used. Meanwhile, the model predicts $S_{\{21,12\}}^{\{\text{TE,TM}\}}=0$ and $S_{\{22\}}^{\{\text{TE,TM}\}}=-1$ (not plotted), and thus the model behaves precisely as a PEC with backwards illumination. 



Finally, note that \eqref{Eq:GroundedSlabChi} satisfies \eqref{Eq:Rotation}. That is, the susceptibility tensors are isotropic---as should be the case for this rotationally symmetric structure---and thus are valid for any polarization or plane of incidence (or any incident field, in general).

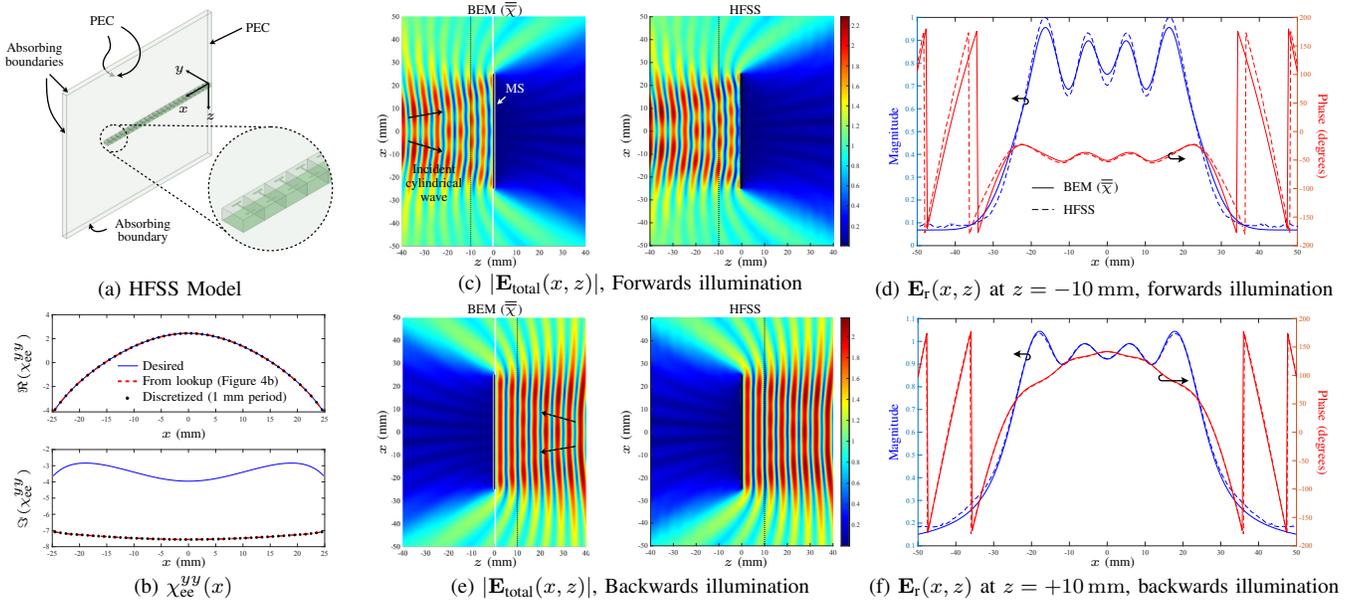
\begin{figure*}
	\begin{overpic}[grid=false,trim=0cm 0cm 0cm 0cm,clip]{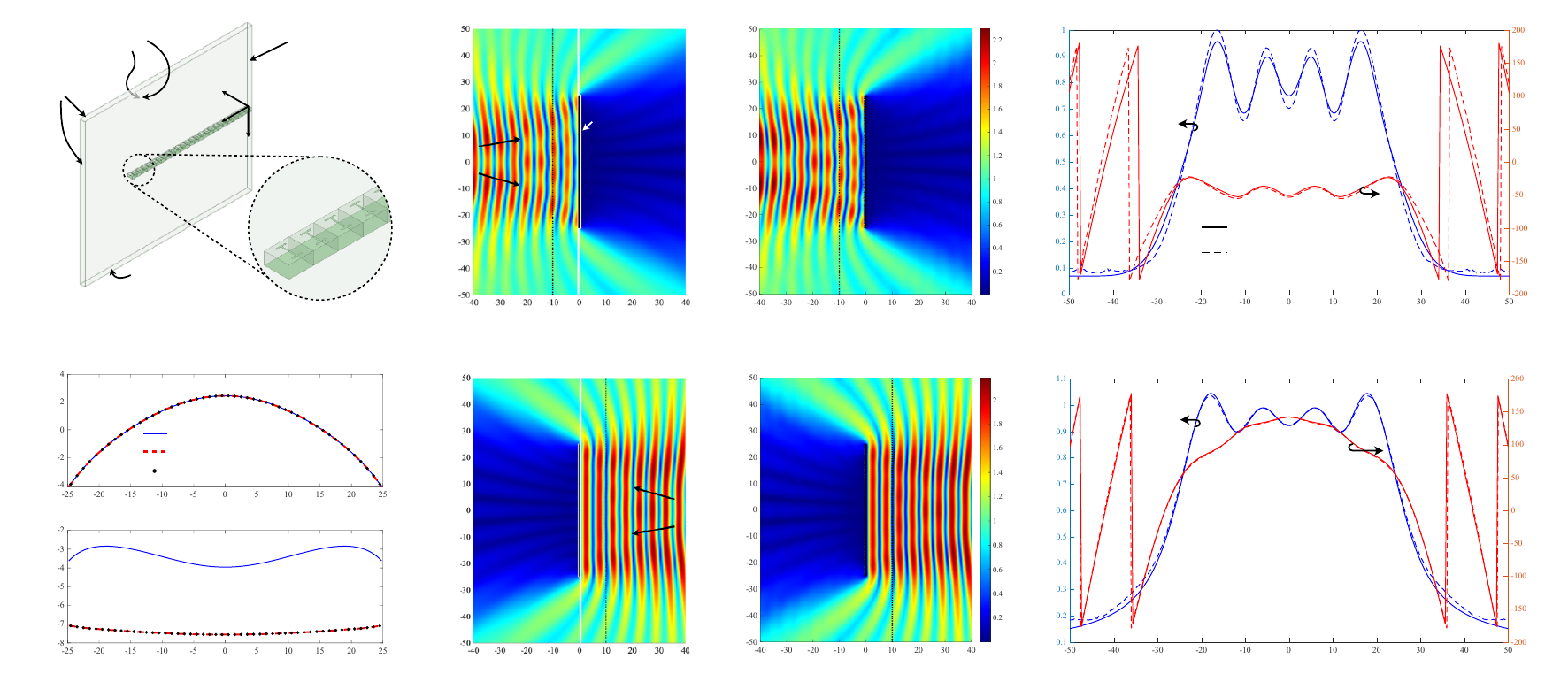}
		\put(13,22){\htext{\footnotesize (a) HFSS Model}}
		\put(14,0){\htext{\footnotesize (b) $\chia{ee}{yy}(x)$}}
		\put(47,22.4){\htext{\footnotesize (c) $|\mathbf{E}_\text{total}(x,z)|$, Forwards illumination}}
		\put(47,0){\htext{\footnotesize (e) $|\mathbf{E}_\text{total}(x,z)|$, Backwards illumination}}
		\put(82,22){\htext{\footnotesize (d) $\mathbf{E}_\text{r}(x,z)$ at $z=\SI{-10}{mm}$, forwards illumination}}
		\put(82,0){\htext{\footnotesize (f) $\mathbf{E}_\text{r}(x,z)$ at $z=+\SI{10}{mm}$, backwards illumination}}
		\put(3,40){\htext{\tiny Absorbing}}
		\put(3,39){\htext{\tiny boundaries}}
		\put(19.5,41.5){\htext{\tiny PEC}}
		\put(8,42){\htext{\tiny PEC}}
		\put(11,27){\htext{\tiny Absorbing}}
		\put(11,26){\htext{\tiny boundary}}
		\put(16,35){\htext{\tiny $z$}}
		\put(14,35.5){\htext{\tiny $x$}}
		\put(13.7,38.5){\htext{\tiny $y$}}
 		\put(11,16.5){\tiny Desired}
		\put(11,15.4){\tiny From lookup (Figure~\ref{Fig:CellExtraction}b)}
		\put(11,14.2){\tiny Discretized (1 mm period)}
		\put(14,2){\htext{\tiny $x$ (mm)}}
		\put(14,11.5){\htext{\tiny $x$ (mm)}}
		\put(2,6.5){\vtext{\tiny $\Im(\chia{ee}{yy})$}}
		\put(2,16.5){\vtext{\tiny $\Re(\chia{ee}{yy})$}}
		\put(37,2){\htext{\tiny $z$ (mm)}}
		\put(55.5,2){\htext{\tiny $z$ (mm)}}
 		\put(28.5,11.4){\vtext{\tiny $x$ (mm)}}
 		\put(46.5,11.4){\vtext{\tiny $x$ (mm)}}
 		\put(37,20.7){\htext{\tiny BEM ($\chiat{}$)}}
		\put(55.5,20.7){\htext{\tiny HFSS}}
		\put(38.5,37){\htext{\tiny \color{white} MS}}
		\put(32.5,31){\htext{\tiny Incident}}
		\put(32.5,30){\htext{\tiny cylindrical}}
		\put(32.5,29){\htext{\tiny wave}}
		\put(37,24.2){\htext{\tiny $z$ (mm)}}
		\put(55.5,24.2){\htext{\tiny $z$ (mm)}}
 		\put(28.5,33.3){\vtext{\tiny $x$ (mm)}}
 		\put(46.5,33.3){\vtext{\tiny $x$ (mm)}}
 		\put(37,42.8){\htext{\tiny BEM ($\chiat{}$)}}
		\put(55.5,42.8){\htext{\tiny HFSS}}
		\put(82.6,2){\htext{\tiny $x$ (mm)}}
		\put(66.3,11.4){\vtext{\tiny \color{blue}Magnitude}}
		\put(98.3,11.4){\vtextb{\tiny \color{red}Phase (degrees)}}
		\put(82.6,24.2){\htext{\tiny $x$ (mm)}}
		\put(66.3,33.3){\vtext{\tiny \color{blue}Magnitude}}
		\put(98.3,33.3){\vtextb{\tiny \color{red}Phase (degrees)}}
		\put(79,29.7){\tiny BEM ($\chiat{}$)}
		\put(79,28){\tiny HFSS}
	\end{overpic}
	\caption{A parabolic reflector was designed using the unit cell from Figure~\ref{Fig:CellExtraction}, having a focal length of \SI{10}{cm} with forwards illumination. (a) HFSS simulation model of half of the structure, with symmetry in the $y-z$ plane, and a total of $2\times 25$ cells; i.e., total length of \SI{5}{cm}. (b) Desired and realized susceptibilities for reflecting a plane wave. (c) Comparison of the \textit{total} field magnitude, using a boundary element method (BEM) which implements the GSTCs and HFSS (full-wave), with an incident cylindrical wave originating from $\mathbf{r}_\text{s}=(\SI{0}{cm},\SI{-10}{cm})$. (d) Comparison of \textit{reflected} fields at $z=\SI{-10}{mm}$. (e-f) Same as (c-d) but with an incident cylindrical wave originating from $\mathbf{r}_\text{s}=(\SI{0}{cm},+\SI{10}{cm})$}\label{Fig:Reflector}
\end{figure*}

\subsection{Sub-wavelength Resonator on a Dielectric Slab} \label{Sec:UnitCell}

Next, we extend the structure by including an array of electric dipoles on top of the grounded slab; this can truly be called a metasurface. For simplicity, we will only consider TE polarization, since the unit cell was designed to resonate with TE polarization. Shown in Figure~\ref{Fig:CellExtraction}, the unit cell is deeply sub-wavelength ($\lambda/10$ at \SI{30}{GHz}). It has a ``dogbone''-shaped copper dipole loaded with a lumped inductance ($L$) at the center, placed on a Rogers RO4003C substrate (\SI{508}{\micro m} thickness; i.e. $kd=0.32$), on top of a PEC ground-plane. 

While there is no analytical expression for $S_{11}^{\text{TE}}(\theta)$, it can be found through full-wave simulations of the unit cell \cite{achouri_general_2015,liu_generalized_2019}. In particular, two angles of incidence need to be simulated. Then, using \eqref{Eq:Sparams} to solve the susceptibilities\footnote{This is similar to Section~\ref{Sec:GroundedSlab}, but with the simulated S-parameters rather than analytical expressions},
\begin{subequations}
	\begin{align}
		\chia{mm}{xx} &= 0\\
        \chia{em}{yx} &= -\frac{2j}{k_0}\\
        \chia{ee}{yy} &=  \frac{4j(\SP{11}{TE}(0)-1)}{k(\SP{11}{TE}(0)+1)}
	\end{align}
	\begin{multline}
	    \chia{mm}{zz} = \frac{4j}{k}\csc\theta\left[\csc\theta\left( \frac{2}{\SP{11}{TE}(0)+1}-1 \right)\right.\\
	    \left. 
	    +\cot\theta\left( 1-\frac{2}{\SP{11}{TE}(\theta)+1} \right)
	    \right]
	\end{multline}
	\label{Eq:UnitCellChi}
\end{subequations}
with all other terms being irrelevant for TE polarization due to the lack of polarization conversion.

This extraction was performed using full-wave simulations (Ansys HFSS) of the unit cell with periodic boundaries, using $\theta\in\{\SI{0}{\degree},\SI{60}{\degree}\}$, and with $\SI{1.2}{nH}\leq L \leq \SI{1.45}{nH}$; the susceptibilities are plotted in Figure~\ref{Fig:CellExtraction}b. Both $\chia{ee}{yy}$ and $\chia{mm}{zz}$ are extracted, while $\chia{em}{yx}$ only depends on frequency through (\ref{Eq:UnitCellChi}b) (not plotted). Subsequently, the predicted reflection is plotted in Figure~\ref{Fig:CellExtraction}c and compared to the full-wave simulated one for $L=\SI{1.4}{nH}$, at which there is a resonance at \SI{30}{GHz}. We observe nearly perfect agreement with the HFSS simulations for $\SP{11}{TE}$ at the two different angles of incidence, while we also have $\SP{22}{TE}=-1$ corresponding to an ideal PEC at all frequencies (not shown here) thanks to the inclusion of $\chia{em}{yx}$. Thus, this combination of $\chia{ee}{yy}$, $\chia{em}{yx}$, and $\chia{mm}{zz}$ produces a very accurate model for scattered TE-polarized fields, regardless of the incident field.


\section{Numerical Demontration: Parabolic Reflector}\label{Sec:Reflector}
Finally, we demonstrate the utility of the two-sided susceptibility model of the reflective unit cell from Section~\ref{Sec:UnitCell} by designing a parabolic reflector; i.e. a metasurface that reflects a plane wave when a cylindrical wave is present from a line source (forward illumination). The incident field produced by the line source is 
\begin{subequations}
\begin{align}
    E_{\text{i},y}(\mathbf{r}) &=  E_0\frac{H_{0}^{(2)}\left(k|\mathbf{r}-\mathbf{r}_\text{s}|\right)}{H_{0}^{(2)}\left(k|\mathbf{r}_\text{s}|\right)} \\
    H_{\text{i},x}(\mathbf{r}) &=E_0\frac{j(z-z_\text{s}) H_{1}^{(2)}\left(k|\mathbf{r}-\mathbf{r}_\text{s}|\right)}{\eta|\mathbf{r}|H_{0}^{(2)}\left(k|\mathbf{r}_\text{s}|\right)}\\
    H_{\text{i},z}(\mathbf{r}) &=-E_0\frac{j(x-x_\text{s}) H_{1}^{(2)}\left(k|\mathbf{r}-\mathbf{r}_\text{s}|\right)}{\eta|\mathbf{r}|H_{0}^{(2)}\left(k|\mathbf{r}_\text{s}|\right)}
\end{align}\label{Eq:ReflectorEi}
\end{subequations}
where $\mathbf{r}_\text{s}=(x_\text{s},z_\text{s})$ is the location of the source, $H_{\{0,1\}}^{(2)}$ are Hankel functions of the second kind, of order 1 and 2, and $E_0$ is the amplitude. For our example, $\mathbf{r}_\text{s}=(\SI{0}{cm},-\SI{10}{cm})$, at 30 GHz. Meanwhile, the reflected field should have a uniform phase, while we taper the amplitude so that the surface remains passive. Specifically,
\begin{subequations}
    \begin{align}
        E_{\text{r},y}(x,0) =0.9\left|E_{\text{i},y}(x,0)\right|\\
        H_{\text{r},x}(x,0) =\frac{E_{\text{r},y}(x,0)}{\eta}
    \end{align}\label{Eq:ReflectorEr}
\end{subequations}

Using \eqref{Eq:ReflectorEi} and \eqref{Eq:ReflectorEr}, we calculate the ideal $\chia{ee}{yy}$ point-wise along $x$ using (\ref{Eq:UnitCellChi}c), treating $\chia{mm}{zz}$ as a perturbation.\footnote{That is, we ignore $\chia{mm}{zz}$ when designing the unit cell distribution, but will subsequently include it in the susceptibility model, for the prediction of fields.} We do not aim to optimize the design, but rather wish to show how the two-sided susceptibility model is accurate for a finite-sized and non-uniform surface. This produces the ``desired'' profile in Figure~\ref{Fig:Reflector}b. However, the unit cell extraction is limited in replicating this profile since there is only one parameter $L$, so that the real and imaginary parts of $\chia{ee}{yy}$ cannot be tuned independently. For the sake of this demonstration, the real part is realized, using the look-up plot from Figure~\ref{Fig:CellExtraction}b. Meanwhile, the $\Im(\chia{ee}{yy})$ (and $\chia{mm}{zz}$) are also included in the susceptibility model. As seen in Figure~\ref{Fig:Reflector}b, the imaginary part deviates from the desired profile. Finally, the profile is discretized into 50 cells, for a length of \SI{50}{mm}.

To check the accuracy of the susceptibility model for this finite non-uniform MS, we use an integral-equation simulator which implements the complete dipolar GSTCs \cite{smy_ie-gstc_2021}. Meanwhile, the structure is simulated in HFSS using the model in Figure~\ref{Fig:Reflector}a, where we leverage symmetry in the $y-z$ plane to reduce the size of the model by half (25 unit cells are simulated). The magnitude of the total fields are compared in Figure~\ref{Fig:Reflector}c, for a cylindrical wave originating from $\mathbf{r}_\text{s}=(\SI{0}{cm},\SI{-10}{cm})$, i.e., forwards illumination. We see a good match between the predicted total fields based on the non-uniform susceptibilities and the full-wave simulation, which takes orders of magnitude longer in time to simulate. Moreover, we see that the susceptibility model correctly models the diffraction around the edges of the finite surface, which is where ``backward illumination'' behavior is critical. The \textit{reflected} fields at the $z=-\SI{10}{mm}$ plane are plotted for closer examination in Figure~\ref{Fig:Reflector}d. While an ideal parabolic reflector would produce a uniform amplitude and phase, there are fluctuations, which may be explained by the finite size of the reflector and edge diffraction, and the fact that the unit cell had only a single parameter of control so that only $\Re(\chia{ee}{yy})$ was implemented. However, it is not our purpose to optimize the design, which would require a more complex unit cell with loss, but rather to show the accuracy of the susceptibility model. To this end, we see an excellent match between BEM and HFSS. 

Lastly to exemplify the two-sided GSTCs, in Figure~\ref{Fig:Reflector}e we repeat the simulations with a cylindrical wave originating from $\mathbf{r}_\text{s}=(\SI{0}{cm},\SI{+10}{cm})$, i.e., backward illumination. As expected, we observe a good match for the \textit{total} fields, but the interference obscures the difference between the forwards and backwards illumination cases. By plotting the \textit{reflected} fields alone (Figure~\ref{Fig:Reflector}e), we see that there is a curved phase, which is to be expected for a PEC sheet with an incident cylindrical wave. Thus, the PEC behavior with backwards illumination is correctly modeled and confirmed.

\section{Conclusions}\label{Sec:Conclusion}

We have shown that it is possible to use the conventional two-sided GSTCs and the HK model for the surface susceptibilities to predict the fields scattered from a fully-reflective metasurface. The validity of the HK model may not be a priori obvious, since this model defines the acting fields as the average of the fields on either side of the metasurface, which seems to be in contradiction with the considered cases where the two sides are in fact electrically isolated and independent. Nevertheless, this approach proved to be in good agreement with full-wave simulations. Indeed, the retrieved susceptibilities from the model work for both forwards and backwards illumination, behaving as a PEC for the latter, which is important if the surface is finite such that the fields may interact with the reverse side. We have also shown the mapping of the geometrical and electrical properties of a dielectric slab (and a PEC sheet with a dielectric cover layer) to susceptibilities, showing the role of the normal and bi-anisotropic terms. These equivalent surface susceptibilities thus act as compact models for these practically volumetric structures and may be easily integrated into a variety of simulation platforms to enable an efficient computation of scattered field from finite-sized volumetric structure for a faster iterative design flow.

\bibliographystyle{IEEEtran}
\bibliography{zotero.bib}

\end{document}